\documentclass{article}

\usepackage{PRIMEarxiv}

\usepackage[utf8]{inputenc}
\usepackage[T1]{fontenc}
\usepackage{hyperref}
\usepackage{url}
\usepackage{booktabs}
\usepackage{amsfonts}
\usepackage{amssymb}
\usepackage{amsthm}
\usepackage{nicefrac}
\usepackage{microtype}
\usepackage{amsmath}
\usepackage{fancyhdr}
\usepackage{graphicx}
\usepackage{xcolor}
\graphicspath{{figs/}{./}}

\title{Two-Step QAOA: Enhancing Quantum Optimization by Decomposing K-hot Constraints in QUBO Formulations}

\author{
  Yuichiro Minato \\
  blueqat Inc. \\
  Shibuya2-24-12, Shibuya \\
  Tokyo, Japan\\
  \texttt{minato@blueqat.com} \\
}

\begin{document}
\maketitle

\begin{abstract}
Penalty-based QUBO formulations force the user of the Quantum Approximate Optimization Algorithm (QAOA) to choose a penalty weight: too large and the cost differences between feasible solutions become small compared with the scale the circuit angles are tuned to, too small and the constraint is violated.
Two-Step QAOA removes this choice for k-hot constraints $\sum_{i\in g}x_i=k$.
A first QAOA stage, driven only by the penalty $(\sum_{i\in g}x_i-k)^2$ and the transverse-field mixer, prepares the initial state; a second stage optimizes the cost alone with a constraint-preserving XY mixer.
We show that (i) stage 2 cannot change the probability of satisfying the hardened constraints, which is therefore fixed by stage 1 and enforced exactly by post-selection;
(ii) for each group the constraint-satisfying component of the stage-1 state is exactly a Dicke state, so stage 1 is a Dicke-state preparation up to a leakage amplitude, and Two-Step QAOA has, as a function of the stage-2 angles, the same post-selected landscape as the Dicke-initialized alternating-operator ansatz;
(iii) the stage-1 angles depend only on the group size and $k$, not on the problem, and can be tabulated once; for half filling, three layers reach a k-hot probability of at least $0.994$ for all tested sizes up to 32 qubits per group.
We present the method as a compilation pass that takes the constraints as they are already written in penalty form.
In state-vector simulations of cardinality-constrained portfolios (60 instances, up to 12 qubits), Two-Step QAOA with a complete-graph XY mixer reaches the optimum more often than penalty QAOA under every penalty-selection rule we tested, at every depth up to five and at the same training budget (median paired ratios $2.9$--$288$ against deployable rules, and $2.2$--$9.7$ at depth $\ge3$ against an oracle-tuned weight given five times the budget); with the cheaper ring mixer the advantage holds at shallow depth but a small penalty weight catches up at depth three to five for half filling.
For graph coloring, where all constraints can be hardened, it finds an optimum with probability $0.93$ at 15 qubits, against $0.13$ for penalty QAOA and $0.12$ for uniform sampling.
For permutation constraints only the rows can be hardened; this raises feasibility and beats penalty QAOA, but stays on par with uniform sampling of permutations for QAP and TSP.
\end{abstract}

\section{Introduction}
\label{sec:intro}
The Quantum Approximate Optimization Algorithm (QAOA)~\cite{farhi2014quantumapproximateoptimizationalgorithm} and its generalization, the Quantum Alternating Operator Ansatz~\cite{Hadfield_2019}, are the standard gate-based approaches to combinatorial optimization on near-term devices.
Most applications~\cite{Suksmono_2022,niroula2022constrained,Tabi_2020} start from a QUBO~\cite{lucas2014ising} in which constraints appear as penalty terms $A(\cdot)^2$.
The penalty weight $A$ must be chosen per problem: a weight large enough to guarantee feasibility of the QUBO minimum compresses the cost differences that QAOA has to resolve, while a small weight lets the optimizer trade violations for cost.
In practice the weight is searched, e.g.\ by starting small and doubling it whenever the returned answer is infeasible, and each step of the search costs a full training run.

For one-hot and, more generally, k-hot constraints, the alternating-operator ansatz offers a way out: an XY mixer preserves the Hamming weight of a group, so the constraint becomes hard, provided the circuit starts in the feasible subspace~\cite{Hadfield_2019,wang2020xy}.
The usual starting point is a Dicke state, prepared by a dedicated circuit~\cite{B_rtschi_2019,baertschi2022short}.

The first version of this work proposed to prepare that starting point with QAOA itself: a first stage driven by the penalty term only, followed by an XY-mixer stage driven by the cost only. This replaces the penalty weight and the Dicke circuit by a second QAOA run. The earlier versions described the idea without analysis or numerical evidence.
This version adds both. It keeps the two-stage construction of the earlier versions (stage 1: transverse-field mixer and the constraint term without a weight; stage 2: XY mixer and the cost), and changes or sharpens the following points:
(a) the stage-1 output was described as ``close to'' the initial state required by the XY mixer; we show that its feasible component is exactly a Dicke state and that the only deviation is a leakage probability $1-P_k$;
(b) the constraint was said to become hard; this holds only together with post-selection of the measured samples, which we add to the method, because stage 2 cannot reduce the leakage;
(c) the angles of both stages were to be found by a classical optimizer; stage 1 is now problem-independent and read from a table, and only stage 2 is trained;
(d) the statement that no penalty weight has to be balanced applies to the constraints that are hardened; constraints that cannot be hardened (overlapping groups, inequalities) keep their penalties in the stage-2 cost;
(e) the general review of adiabatic computation and QAOA is shortened, and the title and equations are corrected (k-hot instead of one-hot, layer-wise angles, an explicit mixer).
Our contributions are:
\begin{enumerate}
\item \textbf{Exact properties} (Sec.~\ref{sec:theory}): stage 2 conserves the probability of satisfying the hardened constraints; the feasible component of the stage-1 state is exactly a (product of) Dicke state(s); Two-Step QAOA and the Dicke-initialized ansatz have the same post-selected landscape; and the k-hot structure of a penalty is not identifiable from a QUBO matrix alone.
\item \textbf{A compilation pass} (Sec.~\ref{sec:software}): the constraints are taken as written in penalty form, stage 1 is looked up in a problem-independent table computed in the symmetric subspace, and no penalty weight is searched for the hardened constraints.
\item \textbf{Numerical evidence} (Sec.~\ref{sec:results}), including deployable penalty-search baselines, uniform feasible sampling, and the cases where the method does not help.
\end{enumerate}

\section{Background}
\label{sec:background}

\subsection{QAOA and the alternating operator ansatz}
For a cost Hamiltonian $H_C$ diagonal in the computational basis, a depth-$p$ QAOA state is
\begin{equation}
|\boldsymbol\gamma,\boldsymbol\beta\rangle=\prod_{l=1}^{p} e^{-i\beta_l H_M}e^{-i\gamma_l H_C}\,|\psi_0\rangle ,
\end{equation}
with layer-wise angles $(\gamma_l,\beta_l)$ chosen by a classical optimizer to minimize $\langle H_C\rangle$.
Standard QAOA uses $H_M=\sum_iX_i$ and $|\psi_0\rangle=|+\rangle^{\otimes n}$.
In the alternating-operator ansatz~\cite{Hadfield_2019} the mixer is chosen to preserve a feasible subspace; for a k-hot group the XY mixer, built from $(X_iX_j+Y_iY_j)/2$ terms on pairs inside the group, preserves the group weight $W_g=\sum_{i\in g}x_i$~\cite{wang2020xy}.
The initial state must then lie in the feasible subspace; it need not be an eigenstate of the mixer.
A natural choice is the Dicke state $|D(n,k)\rangle=\binom{n}{k}^{-1/2}\sum_{|x|=k}|x\rangle$, the uniform superposition of all distinct weight-$k$ strings.
It is an eigenstate of the complete-graph XY Hamiltonian, and of the ring XY Hamiltonian only for $k\in\{0,1,n-1,n\}$; the ordered gate products used in circuits (Eq.~\eqref{eq:xyproduct}) need not have it as an eigenstate.

\subsection{Penalty formulation}
A QUBO with k-hot constraints reads
\begin{equation}
H=C(x)+\sum_{g}A_g\Big(\sum_{i\in g}x_i-k_g\Big)^2 ,
\end{equation}
where a penalty contributes the coupling $2A_g$ to every pair inside $g$ and $A_g(1-2k_g)$ to every linear term.
The weight must be large enough to make violations unattractive, but every unit of $A_g$ shrinks the relative size of the cost differences in $H$.

\section{Two-Step QAOA}
\label{sec:method}
Let $\mathcal G$ be a set of pairwise-disjoint k-hot groups chosen for hardening.
\textbf{Stage 1} starts from $|+\rangle^{\otimes n}$ and applies, on each group $g$ independently, $p_1^{(g)}$ layers of
\begin{equation}
e^{-i\beta^{(1,g)}_l\sum_{i\in g}X_i}\,e^{-i\gamma^{(1,g)}_l (W_g-k_g)^2},
\end{equation}
i.e.\ QAOA on the penalty alone, with no weight to choose. Because the groups are disjoint these circuits commute; a single QAOA on $H_{C_1}=\sum_g(W_g-k_g)^2$ with shared angles is the special case of equal angles and depths.
\textbf{Stage 2} starts from the stage-1 output $|\psi_1\rangle$ and applies $p_2$ layers of
\begin{equation}
U_{XY}(\beta^{(2)}_l)\,e^{-i\gamma^{(2)}_l H_{C_2}},
\end{equation}
where $H_{C_2}$ is the original QUBO with the penalties of the hardened groups removed (penalties of constraints that are not hardened remain), and $U_{XY}$ is the XY mixer of Eq.~\eqref{eq:xyproduct} on the hardened groups (with the $X$ mixer on the remaining qubits). We use two edge sets inside a group: the parity-ordered ring ($|g|$ pairs per layer) and the complete graph ($|g|(|g|-1)/2$ pairs in lexicographic order, the same number of two-qubit terms as a dense cost layer); within a layer the pair rotations are applied in the listed order.
All angles are layer-wise.
The output is measured and samples violating a hardened constraint are discarded.
The earlier versions left open how the angles of the two stages are optimized; here stage 1 is fixed from a table and stage 2 is trained on the post-selected expectation of $H_{C_2}$, which Sec.~\ref{sec:theory} shows to be the natural objective.

\section{Properties of the Two-Step Construction}
\label{sec:theory}

Throughout, $\mathcal{G}=\{g_1,\dots,g_m\}$ denotes the k-hot constraints selected for hardening,
$\sum_{i\in g}x_i=k_g$, and we assume the groups are pairwise disjoint.
Let $W_g=\sum_{i\in g}x_i$ (diagonal in the computational basis) and let
$\Pi=\prod_{g}\Pi_g$, with $\Pi_g$ the projector onto $W_g=k_g$, be the projector onto the
hard-feasible subspace. Stage~1 uses the initial state $|+\rangle^{\otimes n}$, the transverse-field
mixer $B=\sum_i X_i$ and the cost $H_{C_1}=\sum_{g}(W_g-k_g)^2$; stage~2 uses the cost
$H_{C_2}$ (the original QUBO with the penalties of the hardened groups removed) and the mixer
\begin{equation}
U_{XY}(\beta)=\prod_{g\in\mathcal{G}}\ \prod_{(i,j)\in E_g} e^{-i\beta (X_iX_j+Y_iY_j)/2}
\label{eq:xyproduct}
\end{equation}
for a fixed ordered edge list $E_g$ inside each group (the parity-ordered ring or the complete graph in lexicographic order),
together with $e^{-i\beta X_i}$ on qubits that belong to no hardened group.
Equation~\eqref{eq:xyproduct} is an ordered product, not the exponential of the summed XY Hamiltonian;
everything below holds for either choice.

\paragraph{Proposition 1 (stage 2 conserves hard feasibility).}
Every factor of stage~2 commutes with $\Pi$. Hence for any stage-2 depth $p_2$ and any angles,
the probability that the output satisfies all \emph{hardened} constraints is fixed by stage~1,
\begin{equation}
P_{\mathrm{hard}}(\text{final}) = \langle\psi_1|\Pi|\psi_1\rangle \equiv P_{\mathcal{G}} ,
\end{equation}
where $|\psi_1\rangle$ is the stage-1 output. Constraints that are not hardened (kept as penalties in $H_{C_2}$) are not covered: their satisfaction
probability does depend on the stage-2 angles.
\emph{Proof.} $\Pi$ is diagonal in the computational basis, so it commutes with every diagonal operator, in particular with $H_{C_2}$ (whatever penalties of non-hardened constraints it contains). $X_iX_j+Y_iY_j=2(\sigma^+_i\sigma^-_j+\sigma^-_i\sigma^+_j)$ with $i,j\in g$
preserves $W_g$ and does not act on other groups; single-qubit $X$ acts only outside all hardened groups.
\qed

Consequently stage~2 can neither repair nor create violations of the hardened constraints: the infeasible
component produced by stage~1 propagates to the output. A hard-constraint guarantee therefore
requires either $P_{\mathcal{G}}=1$ or post-selection on $\Pi$ (discarding violating samples),
which costs $1/P_{\mathcal{G}}$ shots per kept sample. Disjointness is essential: an XY term on a pair
that straddles two overlapping groups changes both group weights.

\paragraph{Proposition 2 (exact Dicke component).}
For a single group ($m=1$, $g=\{1..n\}$, weight $k$), stage~1 is invariant under all qubit permutations,
so $|\psi_1\rangle$ lies in the $(n+1)$-dimensional symmetric subspace spanned by Dicke states and
\begin{equation}
\Pi_k|\psi_1\rangle = a_k\,|D(n,k)\rangle,\qquad |a_k|^2=P_k .
\end{equation}
For disjoint groups stage~1 acts on each group separately, so $|\psi_1\rangle=\bigotimes_g|\psi_1^{(g)}\rangle\otimes|+\rangle^{\otimes n_{\mathrm{free}}}$
and $\Pi|\psi_1\rangle\propto\bigotimes_g|D(|g|,k_g)\rangle\otimes|+\rangle^{\otimes n_{\mathrm{free}}}$,
with $P_{\mathcal{G}}=\prod_g P_{k_g}^{(g)}$. A per-group target $P^{(g)}\ge 1-\epsilon$ only gives
$P_{\mathcal G}\ge(1-\epsilon)^m$, so a global leakage budget has to be divided among the $m$ groups.

Thus stage~1 is an exact Dicke-state preparation up to a leakage amplitude, not an approximation
of it: the post-selected stage-1 state \emph{is} the Dicke state. Its only figure of merit is $P_k$,
which depends on $(n,k,p_1)$ and the stage-1 angles but \emph{not on the problem}. The stage-1 angles
can therefore be tabulated once and shipped with the software (Sec.~\ref{sec:software}). Because the
dynamics stays in the symmetric subspace, the table is computed with $(n+1)$-dimensional matrices,
for $n$ far beyond what state-vector simulation of the full circuit allows.

\paragraph{Corollary (relation to the Dicke-initialised alternating-operator ansatz).}
Let $\Pi_{\mathrm{opt}}$ project onto the optima of the original problem (all constraints satisfied),
so that $\Pi_{\mathrm{opt}}\le\Pi$, and assume $P_{\mathcal G}>0$.
Let $P^{\mathrm{Dicke}}_{\mathrm{opt}}(\boldsymbol\gamma_2,\boldsymbol\beta_2)$ be the probability of sampling an optimum
when stage~2 is started from the exact product of Dicke states. Then, for the same stage-2 angles,
\begin{equation}
P^{\mathrm{two\text{-}step}}_{\mathrm{opt}} = P_{\mathcal{G}}\;P^{\mathrm{Dicke}}_{\mathrm{opt}},
\end{equation}
and the post-selected expectation of $H_{C_2}$ coincides with the Dicke-initialised one, so both
optimisations have the same landscape. Two-Step QAOA and the Dicke-initialised ansatz differ only
in (i) how the initial state is prepared and (ii) the leakage $1-P_{\mathcal{G}}$.

\paragraph{Remark (what a QUBO matrix does and does not reveal).}
A penalty $A(W_g-k)^2$ contributes the same coupling $2A$ to every pair inside $g$ and $A(1-2k)$ to each
linear term. The support $g$ and the weight $A$ can often be recognised from the matrix, but $k$ cannot:
\begin{equation}
A(W_g-k)^2 = A(W_g-k')^2 + 2A(k'-k)\,W_g + A(k^2-k'^2),
\end{equation}
so any uniform linear cost on $g$ can be traded for a change of $k$. More generally, an unknown quadratic
cost can absorb changes of the support and of $A$ as well, so no part of the decomposition
``cost $+$ penalties'' is identifiable from the matrix without assumptions on the cost. In software, the constraints
should therefore be passed explicitly (as the formulation code already knows them), and matrix-based
detection used only as a suggestion whose whole decomposition (support, $k$ and $A$) is confirmed by the user (Sec.~\ref{sec:detect}).

\section{Two-Step QAOA as a Compilation Pass}
\label{sec:software}

The construction of Sec.~\ref{sec:theory} needs nothing beyond what a user already writes for a
penalty-based QUBO. We therefore present Two-Step QAOA as a compilation pass that turns a
standard penalty formulation into a constraint-preserving circuit without any penalty tuning
for the hardened constraints.

\paragraph{Input.} A cost QUBO $C(x)$ and a list of equality constraints $\sum_{i\in g}x_i=k_g$,
exactly as passed to a penalty builder such as \texttt{add\_onehot} / \texttt{add\_sum\_eq}.
If only a penalised QUBO matrix is available, the supports $g$ and weights $A_g$ can be suggested
by the detector of Sec.~\ref{sec:detect}, but the whole proposed decomposition (supports, $k_g$ and $A_g$) must be confirmed by the user (Remark in Sec.~\ref{sec:theory}).

\paragraph{Pass.} (The experiments of Sec.~\ref{sec:results} implement steps 2--4 for the groups chosen by the
formulation code; the greedy selection of step 1 is implemented in the detector but was not used in the experiments.)
\begin{enumerate}
\item \emph{Select hard groups.} Choose a pairwise-disjoint subset $\mathcal G$ of the constraints
  (we take groups greedily in order of decreasing number of qubits). Overlapping constraints that are not selected
  (e.g.\ the columns of an assignment matrix when its rows are hardened) stay in the cost as penalties.
\item \emph{Stage 1 from a table.} For each $g\in\mathcal G$ look up stage-1 angles for $(|g|,k_g)$
  in a precomputed table (Table~\ref{tab:stage1}); no per-problem optimisation is performed.
  Groups are independent, so different groups may use different depths and angles.
\item \emph{Stage 2.} Build the cost layer from $C$ plus the penalties of the non-hardened constraints,
  and the XY mixer of Eq.~\eqref{eq:xyproduct} on the hardened groups ($X$ mixer elsewhere).
  Train the stage-2 angles on the expectation post-selected on $\Pi$.
\item \emph{Sample and post-select.} Discard samples that violate a hardened constraint; return the best
  remaining sample. By Proposition~1 the returned answer always satisfies the hardened constraints.
\end{enumerate}

\paragraph{What the user no longer does.} (i) Choose or search a penalty weight for the hardened
constraints; (ii) design a Dicke-state circuit; (iii) optimise stage-1 parameters per problem.
What remains: a penalty weight for constraints that cannot be hardened (overlapping or inequality
constraints), and the usual stage-2 training.

\paragraph{The stage-1 table.} Because stage~1 stays in the symmetric subspace, the table is computed
with $(n+1)$-dimensional matrices. We optimise layer by layer (each depth is initialised from the
previous solution by interpolation and by appending a zero layer, plus random starts over the full
periods $\gamma\in[-\pi,\pi]$, $\beta\in[-\pi/2,\pi/2]$), so the tabulated $P_k$ never decreases with depth.

\subsection{Detecting k-hot penalties in a QUBO matrix}
\label{sec:detect}
When only a penalized QUBO matrix is available, we suggest constraints as follows.
(i) Group the positive couplings by value (relative tolerance $10^{-6}$) and keep only values above the largest relative gap of the sorted couplings (penalties are normally chosen larger than the cost).
(ii) For each value, the maximal cliques of the graph of pairs with that coupling are candidate supports with $A=v/2$; in addition, the maximal cliques of the graph of all couplings above the gap are candidates with $A$ equal to half their median coupling (for costs that perturb the penalty couplings); duplicate supports are merged.
(iii) The integers $k_g$ are obtained by a joint least-squares fit of the linear terms, $\ell_i\approx\sum_{g\ni i}A_g(1-2k_g)$, and a candidate is accepted if the fitted $k_g$ is within $0.05$ of an integer in $[1,|g|-1]$.
Enumerating maximal cliques is exponential in the worst case, but the graphs of equal penalty couplings consist of small, mostly disjoint cliques (one per constraint), so the enumeration took at most a few seconds on our largest instances (189 variables). By the Remark in Sec.~\ref{sec:theory} the result is a suggestion, not a deduction.

\section{Numerical experiments}
\label{sec:results}
\subsection{Setup}
\paragraph{Simulation.} Circuits are built with the blueqat SDK (GitHub version 2.2.0) from $R_Z$, $R_{ZZ}$, $R_X$, $R_{XX}$ and $R_{YY}$ gates.
For the parameter sweeps we evaluate the same circuits with an equivalent dense state-vector routine (diagonal phases, single-qubit
rotations and two-level XY rotations). This routine agrees with the blueqat circuits to a state fidelity of $1$ within $10^{-9}$ for all
three ans\"atze and both XY edge sets on random angles at the size of the experiments ($n=10$), and for the ring mixer on the optimized angles of a preliminary
experiment ($n=8$); unit tests compare the building blocks with exact matrix exponentials. Stage-1 tables are computed in the $(n+1)$-dimensional symmetric subspace, verified against the full
$2^n$-dimensional evolution to $10^{-12}$ for $n\le9$. All probabilities, including the training objective, are computed exactly from the state vector; shots enter only through the cost accounting below and the final extraction.

\paragraph{Training.} Every method is trained with the same procedure: the Hamiltonian is divided by its largest absolute entry,
depths are grown one layer at a time (each depth is initialized by interpolating the previous solution~\cite{zhou2020qaoa}, by the
previous solution with an appended zero layer, and by two random starts; depth one uses four random starts; random angles are uniform in $[-0.5,0.5]$), and each start is run for 150 Adam steps keeping the best
iterate. Constraint-preserving methods are trained on the expectation post-selected on the hardened constraints.

\paragraph{Cost accounting.} One unit of sample cost is one measured bit string. Although the objective is evaluated exactly, a training evaluation is charged $1000$ shots and a
gradient two evaluations per angle. This is an optimizer-budget proxy used to compare methods on equal terms, not a physical shot count: the layer angles control many gates whose generators have several spectral gaps, so a valid parameter-shift rule needs more evaluations. Post-selected
training is charged $1/P_{\mathcal G}$ per evaluation. Every final extraction draws $1000$ shots and returns the best feasible sample;
for penalty ladders the shots that decided to stop are the returned answer, and every rung's training and sampling is charged.

\paragraph{Baselines.} \emph{Penalty QAOA} uses the full QUBO with the $X$ mixer. Its weight is chosen by
(a) the ladder of the formulation library we take the problems from (an in-house library of QUBO formulations, snapshot of 2026-10-07): start at $A_{\rm start}$, double while the returned sample is infeasible, up to a
weight $A_{\rm safe}$ that guarantees a feasible QUBO minimum. For the portfolio, $A_{\rm start}=2\max_i\big(|\ell_i|+\sum_{j\ne i}|q_{ij}|\big)$ and $A_{\rm safe}=\sum_i|\ell_i|+\sum_{i<j}|q_{ij}|+1$ where $\ell_i$ and $q_{ij}$ are the coefficients of $x_i$ and $x_ix_j$ ($i<j$) in the cost; the other problems use the library's own bounds,
(b) the same ladder from $A_{\rm start}/16$,
(c) a grid $A\in\sigma\{0.03,0.1,0.3,1,3\}$, where $\sigma$ is the range of the cost over feasible solutions (obtained by enumeration, so this baseline is partially oracle-assisted), selected by the lowest cost among the feasible samples of $1000$ shots per grid point, ties broken by the larger number of feasible samples, and
(d) the same grid selected by the exact $P_{\rm opt}$ (an oracle).
Grids are charged all five training runs. We also report \emph{uniform sampling of feasible solutions} (a classical baseline),
the \emph{biased product state} followed by XY layers and post-selection, and the \emph{Dicke-initialized} XY ansatz.

\paragraph{Metrics.} $P_{\rm feas}$ and $P_{\rm opt}$ are the probabilities that one sample satisfies all constraints and that it is an
optimum of the constrained problem; $P_{\rm opt|feas}=P_{\rm opt}/P_{\rm feas}$. We compare $P_{\rm opt}$, which determines the number of shots needed to see the optimum ($\approx 3/P_{\rm opt}$ for 95\% confidence), through per-instance ratios; the probability that a batch of $1000$ shots contains the optimum, $1-(1-P_{\rm opt})^{1000}$, follows from it.

\subsection{Stage-1 tables}
\begin{table}[t]\centering\small
\caption{Stage-1 table: best found $P_k$ per group size $n$, weight $k$ and depth $p_1$ (symmetric-subspace optimization; lower bounds on the achievable value). $|+\rangle$ and ``product'' are the $P_k$ of $|+\rangle^{\otimes n}$ and of the biased product state. Last column: smallest $p_1$ with $P_k\ge0.99$ (--: not reached by $p_1=6$).}
\label{tab:stage1}
\begin{tabular}{rrrrrrrrrrr}\toprule
$n$ & $k$ & $|+\rangle$ & product & $p_1{=}1$ & 2 & 3 & 4 & 5 & 6 & $p_1^{0.99}$\\\midrule
4 & 1 & 0.250 & 0.422 & 0.6815 & 0.9814 & 1.0000 & 1.0000 & 1.0000 & 1.0000 & 3\\
4 & 2 & 0.375 & 0.375 & 0.8415 & 1.0000 & 1.0000 & 1.0000 & 1.0000 & 1.0000 & 2\\
5 & 1 & 0.156 & 0.410 & 0.6048 & 0.9748 & 0.9985 & 1.0000 & 1.0000 & 1.0000 & 3\\
5 & 2 & 0.312 & 0.346 & 0.7710 & 0.9699 & 0.9985 & 1.0000 & 1.0000 & 1.0000 & 3\\
6 & 1 & 0.094 & 0.402 & 0.5661 & 0.9667 & 0.9919 & 0.9999 & 1.0000 & 1.0000 & 3\\
6 & 2 & 0.234 & 0.329 & 0.6705 & 0.9502 & 0.9939 & 0.9994 & 1.0000 & 1.0000 & 3\\
6 & 3 & 0.312 & 0.312 & 0.7899 & 0.9963 & 1.0000 & 1.0000 & 1.0000 & 1.0000 & 2\\
8 & 1 & 0.031 & 0.393 & 0.5333 & 0.8992 & 0.9915 & 0.9971 & 0.9972 & 1.0000 & 3\\
8 & 2 & 0.109 & 0.311 & 0.5312 & 0.8910 & 0.9844 & 0.9920 & 0.9991 & 0.9996 & 4\\
8 & 4 & 0.273 & 0.273 & 0.7568 & 0.9928 & 0.9999 & 1.0000 & 1.0000 & 1.0000 & 2\\
10 & 1 & 0.010 & 0.387 & 0.5185 & 0.8491 & 0.9625 & 0.9960 & 0.9960 & 1.0000 & 4\\
10 & 2 & 0.044 & 0.302 & 0.4794 & 0.8570 & 0.9730 & 0.9909 & 0.9960 & 0.9979 & 4\\
10 & 5 & 0.246 & 0.246 & 0.7325 & 0.9871 & 0.9997 & 1.0000 & 1.0000 & 1.0000 & 3\\
12 & 1 & 0.003 & 0.384 & 0.5096 & 0.8352 & 0.9551 & 0.9850 & 1.0000 & 1.0000 & 5\\
12 & 3 & 0.054 & 0.258 & 0.4590 & 0.8204 & 0.9692 & 0.9905 & 0.9990 & 0.9990 & 4\\
12 & 6 & 0.226 & 0.226 & 0.7134 & 0.9888 & 0.9987 & 1.0000 & 1.0000 & 1.0000 & 3\\
16 & 1 & 0.000 & 0.380 & 0.4992 & 0.8880 & 0.9359 & 0.9529 & 0.9762 & 0.9886 & --\\
16 & 4 & 0.028 & 0.225 & 0.4146 & 0.8274 & 0.9140 & 0.9688 & 0.9688 & 0.9861 & --\\
16 & 8 & 0.196 & 0.196 & 0.6844 & 0.9860 & 0.9991 & 0.9999 & 1.0000 & 1.0000 & 3\\
20 & 1 & 0.000 & 0.377 & 0.4931 & 0.7734 & 0.9278 & 0.9363 & 0.9921 & 0.9921 & 5\\
20 & 5 & 0.015 & 0.202 & 0.3829 & 0.7946 & 0.9787 & 0.9787 & 0.9787 & 0.9855 & --\\
20 & 10 & 0.176 & 0.176 & 0.6627 & 0.9838 & 0.9988 & 0.9988 & 0.9988 & 0.9988 & 3\\
24 & 1 & 0.000 & 0.376 & 0.4892 & 0.8362 & 0.9364 & 0.9364 & 0.9749 & 0.9749 & --\\
24 & 6 & 0.008 & 0.185 & 0.3584 & 0.6117 & 0.9266 & 0.9266 & 0.9975 & 0.9981 & 5\\
24 & 12 & 0.161 & 0.161 & 0.6455 & 0.9608 & 0.9960 & 0.9960 & 1.0000 & 1.0000 & 3\\
32 & 1 & 0.000 & 0.374 & 0.4844 & 0.7805 & 0.7805 & 0.8781 & 0.8781 & 0.9932 & 6\\
32 & 8 & 0.002 & 0.161 & 0.3218 & 0.7426 & 0.8027 & 0.8027 & 0.8027 & 0.9785 & --\\
32 & 16 & 0.140 & 0.140 & 0.6191 & 0.9493 & 0.9945 & 0.9995 & 0.9995 & 0.9995 & 3\\
\bottomrule\end{tabular}\end{table}

\begin{figure}[t]\centering
\includegraphics[width=\linewidth]{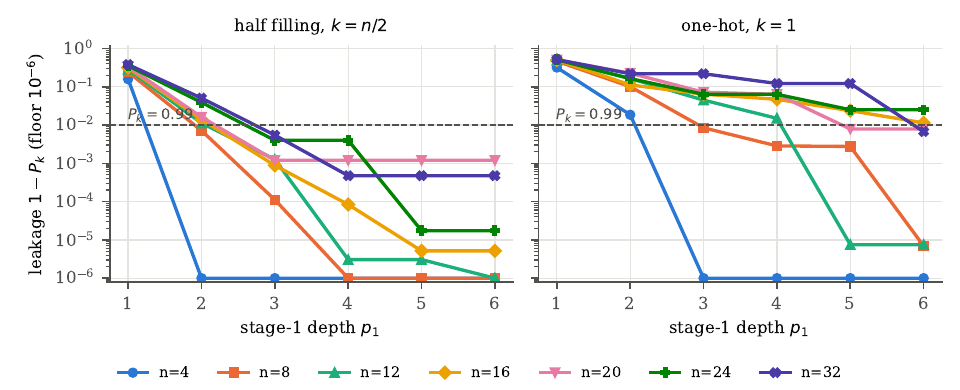}
\caption{Leakage $1-P_k$ of the tabulated stage-1 states versus depth. Left: half filling; right: one-hot.}
\label{fig:stage1}
\end{figure}
Table~\ref{tab:stage1} and Fig.~\ref{fig:stage1} list the stage-1 states for group sizes up to $n=32$.
For half filling, three layers reach $P_k\ge0.9945$ for every tested size ($n=4$--$32$; the lowest value is $0.99454$ at $n=32$), and four layers reach $P_k\ge0.996$; within the tested range, three layers are an observed sufficient depth independent of $n$.
For one-hot groups the smallest depth found reaching $P_k\ge0.99$ grows: three layers up to $n=8$, four for $n=10$, five for $n=12$ and $n=20$, and six for $n=32$, while $n=16$ and $n=24$ stay at $0.989$ and $0.975$ at $p_1=6$. These are values found by local optimization, not lower limits on the depth.
Plateaus such as $n=24$, $k=1$ at $p_1=3$ and $4$ show that the optimizer stopped in a local optimum, so these entries are lower bounds.
For comparison, the entangling-free biased product state has $P_k\approx0.37$--$0.42\approx1/e$ for $k=1$ and $\binom{n}{n/2}2^{-n}$ ($0.14$--$0.27$ for $n\ge8$) for half filling.

\subsection{Cardinality-constrained portfolios}
\begin{figure}[t]\centering
\includegraphics[width=\linewidth]{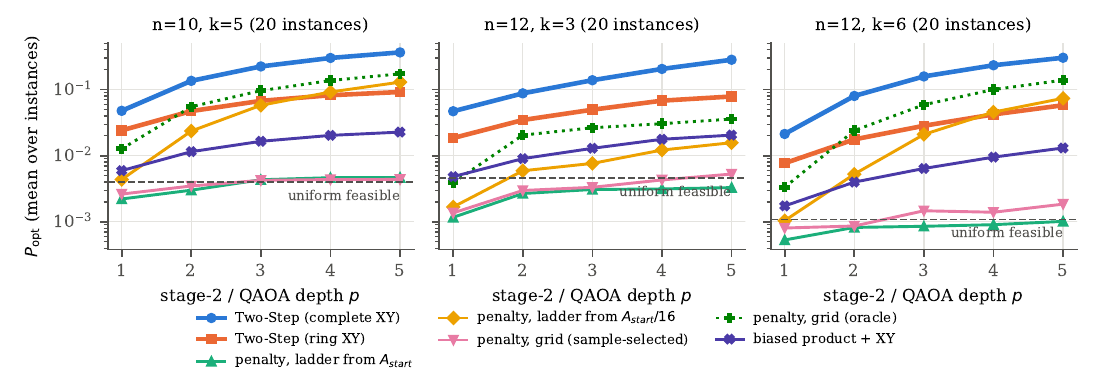}
\caption{Mean optimum probability versus depth for portfolio problems. The Dicke-initialized ansatz agrees with ring-mixer Two-Step up to the factor $P_{\mathcal G}$ (apart from rare differences in the local optimum reached by training) and is not drawn.}
\label{fig:portfolio}
\end{figure}
\begin{table}[t]\centering\small
\caption{Portfolio problems, Two-Step with the ring XY mixer: paired ratio $P_{\rm opt}^{\rm Two\text{-}Step}/P_{\rm opt}^{\rm baseline}$ per instance at the same depth $p$; median with 95\% bootstrap interval of the median, and the number of instances in which Two-Step is better. Budgets: ladders as used (all stopped at their first rung), grids five weights, post-selected methods with overhead $1/P_k$. The biased-product baseline uses the ring XY mixer.}
\label{tab:portfolio}
\resizebox{\linewidth}{!}{\begin{tabular}{llccc}\toprule
$(n,k)$ & baseline & $p=1$ & $p=3$ & $p=5$\\\midrule
(10,5) & uniform feasible & 5.41 [3.46, 7.03] (20/20) & 14.12 [7.96, 18.90] (19/20) & 16.26 [9.53, 28.91] (19/20)\\
(10,5) & penalty, ladder from $A_{\rm start}$ & 9.50 [6.21, 12.66] (20/20) & 12.86 [7.22, 17.73] (19/20) & 14.22 [8.21, 24.41] (19/20)\\
(10,5) & penalty, ladder from $A_{\rm start}/16$ & 5.28 [3.49, 7.10] (20/20) & 1.21 [0.60, 1.73] (11/20) & 0.67 [0.23, 0.81] (5/20)\\
(10,5) & penalty, grid (sample-selected) & 7.50 [5.97, 11.95] (20/20) & 13.23 [7.19, 17.78] (19/20) & 15.26 [8.72, 25.94] (19/20)\\
(10,5) & penalty, grid (oracle-selected) & 1.69 [1.17, 2.29] (17/20) & 0.70 [0.33, 1.20] (7/20) & 0.52 [0.18, 0.82] (4/20)\\
(10,5) & biased product + XY & 4.06 [4.06, 4.06] (20/20) & 4.06 [4.06, 4.06] (20/20) & 4.06 [4.06, 4.06] (20/20)\\
\midrule
(12,3) & uniform feasible & 3.55 [3.28, 4.69] (20/20) & 8.12 [6.58, 11.29] (20/20) & 15.03 [6.43, 22.41] (19/20)\\
(12,3) & penalty, ladder from $A_{\rm start}$ & 13.70 [12.68, 18.20] (20/20) & 12.04 [9.65, 16.70] (20/20) & 20.68 [8.90, 31.13] (19/20)\\
(12,3) & penalty, ladder from $A_{\rm start}/16$ & 9.72 [8.48, 12.79] (20/20) & 5.13 [3.74, 7.46] (19/20) & 4.68 [1.91, 5.99] (17/20)\\
(12,3) & penalty, grid (sample-selected) & 12.81 [10.24, 17.76] (20/20) & 11.06 [9.14, 16.94] (20/20) & 13.92 [5.36, 24.75] (19/20)\\
(12,3) & penalty, grid (oracle-selected) & 5.27 [3.92, 5.85] (20/20) & 2.21 [1.18, 2.81] (15/20) & 1.62 [0.79, 3.21] (12/20)\\
(12,3) & biased product + XY & 3.84 [3.82, 3.84] (20/20) & 3.84 [3.84, 3.84] (20/20) & 3.84 [3.84, 3.84] (20/20)\\
\midrule
(12,6) & uniform feasible & 5.62 [3.94, 8.43] (20/20) & 15.89 [12.22, 30.23] (20/20) & 36.11 [8.78, 71.07] (20/20)\\
(12,6) & penalty, ladder from $A_{\rm start}$ & 11.36 [7.99, 17.11] (20/20) & 20.10 [15.40, 38.47] (20/20) & 39.13 [9.72, 73.92] (20/20)\\
(12,6) & penalty, ladder from $A_{\rm start}/16$ & 5.78 [3.98, 9.10] (20/20) & 1.19 [0.74, 2.63] (13/20) & 0.94 [0.30, 1.62] (8/20)\\
(12,6) & penalty, grid (sample-selected) & 8.04 [5.22, 14.29] (20/20) & 17.49 [8.92, 37.49] (20/20) & 24.92 [7.93, 41.95] (19/20)\\
(12,6) & penalty, grid (oracle-selected) & 1.79 [1.42, 2.69] (18/20) & 0.51 [0.41, 0.80] (6/20) & 0.36 [0.19, 0.55] (3/20)\\
(12,6) & biased product + XY & 4.43 [4.42, 4.43] (20/20) & 4.43 [4.43, 4.43] (20/20) & 4.43 [4.43, 4.43] (20/20)\\
\bottomrule\end{tabular}}\end{table}

We use cardinality-constrained portfolios, $\min_x\,\tfrac12x^\top\Sigma x-\mu^\top x$ subject to $\sum_ix_i=k$, with $\Sigma=AA^\top/n$ for a standard normal $A$ and $\mu_i\sim U(0,1)$, for $(n,k)=(10,5)$, $(12,3)$ and $(12,6)$ and 20 random instances each (Fig.~\ref{fig:portfolio}, Table~\ref{tab:portfolio}).
Three points stand out.

\emph{At shallow depth Two-Step QAOA has the higher optimum probability.} At $p=1$ the median paired ratio is $3.5$--$5.6$ against uniform feasible sampling and $5.3$--$13.7$ against the ladder- and sample-selected penalty rules, in all 20 instances of every size; the oracle-tuned grid with five times the budget is behind by a median factor $1.7$--$5.3$ (Two-Step is better in 17, 20 and 18 of 20 instances).
The ladder that starts from the formulation code's $A_{\rm start}$ always stops at its first rung, because a large weight produces feasible samples immediately; its mean optimum probability then stays at or below the level of uniform sampling ($0.3$--$1.2$ times) at all depths. Selecting the weight from the grid by the sampled answers does not help either: it picks $A\ge0.3\sigma$ in 281 of the 300 (instance, depth) cases, where the optimum probability is low, presumably because batches from such weights already contain good feasible answers.

\emph{With the ring mixer, a small penalty weight catches up at larger depth for half filling.} The ladder started from $A_{\rm start}/16$ (which also stopped at its first rung) improves quickly with depth: at $p=5$ it beats ring-mixer Two-Step in 15 of 20 instances for $(10,5)$ and 12 of 20 for $(12,6)$ (median ratios $0.67$ and $0.94$), at the same training budget, while it remains behind for $(12,3)$ (median $4.7$, 17 of 20). The oracle-tuned grid is better than ring-mixer Two-Step from $p=3$ on for half filling. A small weight leaves roughly half of the samples infeasible ($P_{\rm feas}\approx0.5$--$0.6$), but the $X$ mixer can move through infeasible strings, which the ring XY mixer cannot.

\emph{With the complete-graph mixer, Two-Step is ahead at every depth.} Replacing the ring by all pairs of the group (the same number of two-qubit terms per layer as the cost layer) raises the optimum probability of Two-Step by a median factor $1.4$--$6.3$ (Table~\ref{tab:portfolio_complete}). In median it is then better than every baseline at every depth, including the oracle-tuned grid with five times the budget (median ratios $2.2$--$9.7$ at $p\ge3$, in 18--20 of 20 instances), and better than the small-weight ladder by median factors $2.9$--$17$ at $p=5$. The crossover in the previous paragraph was therefore a limitation of the ring mixer, not of the hard-constraint approach.
\begin{table}[t]\centering\small
\caption{Portfolio problems, Two-Step with the complete-graph XY mixer: paired ratio $P_{\rm opt}^{\rm Two\text{-}Step}/P_{\rm opt}^{\rm baseline}$ per instance at the same depth $p$; median with 95\% bootstrap interval of the median, and the number of instances in which Two-Step is better. Budgets: ladders as used (all stopped at their first rung), grids five weights, post-selected methods with overhead $1/P_k$. The biased-product baseline uses the ring XY mixer.}
\label{tab:portfolio_complete}
\resizebox{\linewidth}{!}{\begin{tabular}{llccc}\toprule
$(n,k)$ & baseline & $p=1$ & $p=3$ & $p=5$\\\midrule
(10,5) & uniform feasible & 10.96 [9.90, 15.72] (19/20) & 50.66 [36.80, 65.18] (20/20) & 87.42 [68.35, 102.58] (20/20)\\
(10,5) & penalty, ladder from $A_{\rm start}$ & 19.25 [17.96, 27.81] (20/20) & 47.12 [32.03, 58.16] (20/20) & 73.32 [59.48, 90.70] (20/20)\\
(10,5) & penalty, ladder from $A_{\rm start}/16$ & 10.97 [8.33, 14.20] (20/20) & 3.68 [3.37, 4.52] (20/20) & 2.87 [2.43, 3.15] (20/20)\\
(10,5) & penalty, grid (sample-selected) & 18.63 [13.82, 25.62] (20/20) & 47.01 [32.17, 64.11] (20/20) & 79.75 [59.32, 96.82] (20/20)\\
(10,5) & penalty, grid (oracle-selected) & 3.60 [2.88, 4.48] (19/20) & 2.21 [1.91, 3.04] (20/20) & 2.24 [1.78, 2.77] (20/20)\\
(10,5) & biased product + XY & 9.15 [5.98, 12.89] (20/20) & 16.65 [10.48, 25.19] (20/20) & 18.34 [12.85, 39.68] (20/20)\\
(10,5) & Two-Step (ring XY) & 2.25 [1.48, 3.18] (17/20) & 4.10 [2.42, 6.20] (19/20) & 4.87 [3.16, 9.77] (19/20)\\
\midrule
(12,3) & uniform feasible & 6.25 [2.31, 7.25] (15/20) & 29.24 [21.55, 36.95] (19/20) & 56.28 [46.33, 66.70] (20/20)\\
(12,3) & penalty, ladder from $A_{\rm start}$ & 24.56 [9.04, 28.73] (17/20) & 42.69 [31.59, 54.51] (19/20) & 75.55 [63.70, 93.84] (20/20)\\
(12,3) & penalty, ladder from $A_{\rm start}/16$ & 17.88 [6.69, 19.19] (16/20) & 17.57 [13.11, 22.09] (18/20) & 17.22 [12.20, 24.27] (19/20)\\
(12,3) & penalty, grid (sample-selected) & 20.13 [8.78, 28.14] (17/20) & 41.68 [30.27, 52.26] (19/20) & 59.67 [44.79, 77.15] (20/20)\\
(12,3) & penalty, grid (oracle-selected) & 7.22 [2.69, 9.56] (15/20) & 6.86 [3.80, 10.50] (18/20) & 9.68 [5.94, 10.56] (19/20)\\
(12,3) & biased product + XY & 5.20 [2.51, 7.15] (15/20) & 9.88 [5.95, 22.05] (19/20) & 17.86 [11.73, 28.25] (20/20)\\
(12,3) & Two-Step (ring XY) & 1.36 [0.65, 1.86] (13/20) & 2.57 [1.55, 5.75] (17/20) & 4.65 [3.06, 7.36] (18/20)\\
\midrule
(12,6) & uniform feasible & 17.49 [13.49, 24.95] (19/20) & 127.53 [84.99, 162.24] (20/20) & 269.61 [163.83, 348.12] (20/20)\\
(12,6) & penalty, ladder from $A_{\rm start}$ & 36.15 [27.19, 51.67] (20/20) & 158.90 [108.86, 205.59] (20/20) & 287.54 [177.34, 370.46] (20/20)\\
(12,6) & penalty, ladder from $A_{\rm start}/16$ & 18.61 [14.74, 24.39] (19/20) & 7.40 [6.04, 8.83] (20/20) & 4.72 [3.63, 7.22] (20/20)\\
(12,6) & penalty, grid (sample-selected) & 24.81 [21.48, 32.33] (20/20) & 152.99 [84.33, 198.96] (20/20) & 167.66 [112.08, 266.49] (20/20)\\
(12,6) & penalty, grid (oracle-selected) & 6.09 [5.22, 7.61] (19/20) & 2.70 [2.29, 4.91] (20/20) & 2.50 [2.19, 2.80] (20/20)\\
(12,6) & biased product + XY & 13.19 [9.24, 21.75] (19/20) & 24.64 [19.03, 49.09] (20/20) & 28.02 [19.01, 51.85] (20/20)\\
(12,6) & Two-Step (ring XY) & 3.00 [2.06, 4.92] (18/20) & 5.56 [4.30, 11.09] (20/20) & 6.33 [4.29, 11.71] (19/20)\\
\bottomrule\end{tabular}}\end{table}

Two further observations confirm the theory. The Corollary predicts $P_{\rm opt}^{\rm Two\text{-}Step}=P_{\mathcal G}P_{\rm opt}^{\rm Dicke}$ for identical stage-2 angles, and, applied to the biased product state (whose weight-$k$ component is also the Dicke state), a fixed ratio $P_{\mathcal G}/P_k^{\rm product}=4.06$, $3.84$ and $4.43$ for the three sizes. Since the three methods are trained from independent starts, the angles need not coincide; in practice both relations hold to within $0.1\%$ in 84--89\% of the (instance, depth) pairs, and the remaining pairs deviate by at most $3\%$ except one pair with $34\%$, where training reached a different local optimum. The median ratios for the ring-mixer biased product state in Table~\ref{tab:portfolio} equal the predicted values.

\subsection{Detecting constraints}
\begin{table}[t]\centering\small
\caption{Detecting k-hot penalties from QUBO matrices of 168 instances of 29 formulations. Truth: the constraints passed to the penalty builders. Final detections are counted correct only if both the support and $k$ match. Totals: 1179 constraints, 562 detections, 487 correct (precision 0.87, recall 0.41); candidate supports before the $k$ fit: 20541, of which 983 match a true support.}
\label{tab:detect}
\begin{tabular}{lrrrr}\toprule
formulation & constraints & detections & correct & wrong\\\midrule
assignment & 72 & 71 & 71 & 0\\
bin-packing & 51 & 0 & 0 & 0\\
clique & 0 & 14 & 0 & 14\\
clustering & 60 & 44 & 44 & 0\\
coloring & 51 & 51 & 51 & 0\\
community & 60 & 60 & 60 & 0\\
facility & 42 & 0 & 0 & 0\\
feature-selection & 6 & 0 & 0 & 0\\
flowshop & 48 & 42 & 42 & 0\\
graph-partition & 6 & 2 & 0 & 2\\
jobshop & 46 & 42 & 42 & 0\\
knapsack & 0 & 2 & 0 & 2\\
maxcut & 0 & 15 & 0 & 15\\
maze & 0 & 7 & 0 & 7\\
mis & 0 & 5 & 0 & 5\\
number-partition & 0 & 7 & 0 & 7\\
portfolio & 6 & 0 & 0 & 0\\
portfolio-constrained & 6 & 0 & 0 & 0\\
production & 30 & 0 & 0 & 0\\
qap & 60 & 60 & 60 & 0\\
qubo & 0 & 1 & 0 & 1\\
set-cover & 0 & 3 & 0 & 3\\
shift & 336 & 0 & 0 & 0\\
sudoku & 71 & 8 & 3 & 5\\
timetable & 30 & 24 & 24 & 0\\
tsp & 90 & 90 & 90 & 0\\
vertex-cover & 0 & 8 & 0 & 8\\
vrp & 108 & 6 & 0 & 6\\
\bottomrule\end{tabular}\end{table}

We applied the detector of Sec.~\ref{sec:detect} to the QUBO matrices produced by the penalty builders of a library of 29 formulations (Table~\ref{tab:detect}); the truth is the list of constraints those builders were given.
Of $562$ detections, $487$ are correct in both support and $k$ (precision $0.87$), but only $41\%$ of the $1179$ constraints are found.
The support is usually recognized (candidate supports cover $983$ true supports), and the losses are where the Remark of Sec.~\ref{sec:theory} predicts them: when the cost has large linear terms on a group (facility location, shift scheduling, vehicle routing), $k$ is misread; when the cost adds to the couplings inside the group (portfolio) or inequality penalties dominate (bin packing, production), the support is missed; and unconstrained problems with uniform couplings (MaxCut, vertex cover) produce false constraints.
The detector is therefore usable as a suggestion for formulations whose penalties dominate uniformly (assignment, coloring, QAP, TSP), but the constraints should be passed explicitly.

\subsection{Several groups: coloring and permutation problems}
\begin{table}[t]\centering\small
\caption{Problems with several one-hot groups (mean over instances, depth $p=5$, ring XY mixer). ``Hard'' rows: the one-hot groups chosen by the formulation code are hardened with XY mixers; for coloring these are all constraints, for the permutation problems only the rows (columns remain penalties). $P_{\rm opt|feas}$ is the optimum probability among feasible samples. Penalty weights follow the formulation code's ladder.}
\label{tab:multi}
\begin{tabular}{lrrlccc}\toprule
problem & qubits & inst. & method & $P_{\rm feas}$ & $P_{\rm opt}$ & $P_{\rm opt|feas}$\\\midrule
assignment 4 & 16 & 5 & uniform feasible & 1.000 & 0.0417 & 0.0417\\
assignment 4 & 16 & 5 & penalty QAOA & 0.063 & 0.0055 & 0.0885\\
assignment 4 & 16 & 5 & Two-Step (rows hard) & 0.614 & 0.1140 & 0.1850\\
\midrule
graph 3-coloring (dense) 4 & 12 & 10 & uniform feasible & 1.000 & 0.3111 & 0.3111\\
graph 3-coloring (dense) 4 & 12 & 10 & penalty QAOA & 0.972 & 0.4043 & 0.4174\\
graph 3-coloring (dense) 4 & 12 & 10 & Two-Step (all hard) & 1.000 & 0.9861 & 0.9861\\
\midrule
graph 3-coloring (dense) 5 & 15 & 10 & uniform feasible & 1.000 & 0.1185 & 0.1185\\
graph 3-coloring (dense) 5 & 15 & 10 & penalty QAOA & 0.788 & 0.1290 & 0.1761\\
graph 3-coloring (dense) 5 & 15 & 10 & Two-Step (all hard) & 1.000 & 0.9302 & 0.9302\\
\midrule
QAP 4 & 16 & 5 & uniform feasible & 1.000 & 0.0583 & 0.0583\\
QAP 4 & 16 & 5 & penalty QAOA & 0.057 & 0.0044 & 0.0800\\
QAP 4 & 16 & 5 & Two-Step (rows hard) & 0.780 & 0.0633 & 0.0822\\
\midrule
TSP 4 & 16 & 5 & uniform feasible & 1.000 & 0.3333 & 0.3333\\
TSP 4 & 16 & 5 & penalty QAOA & 0.025 & 0.0096 & 0.3876\\
TSP 4 & 16 & 5 & Two-Step (rows hard) & 0.848 & 0.3298 & 0.3890\\
\bottomrule\end{tabular}\end{table}

Table~\ref{tab:multi} uses problems from the same formulation library, with the library's own penalty ladders; Two-Step uses the ring mixer here.
\emph{Graph 3-coloring} (dense random graphs with edge probability $0.7$; minimizing the number of monochromatic edges) has one one-hot group per vertex, all disjoint, so every constraint is hardened and no penalty weight remains.
Two-Step then finds an optimal coloring with probability $0.93$ (5 vertices, 15 qubits) and $0.99$ (4 vertices, 12 qubits) at $p=5$, against $0.13$ and $0.40$ for penalty QAOA and $0.12$ and $0.31$ for uniform sampling. (The 12-qubit value is close to saturation, so the 15-qubit case is the informative one.)
\emph{Permutation problems} (assignment, QAP, TSP on $4\times4$ variables) have one-hot rows and columns; only the rows are hardened and the columns remain penalties. Hardening raises $P_{\rm feas}$ from $0.025$--$0.063$ to $0.61$--$0.85$ and improves the optimum probability over penalty QAOA by a factor $14$--$34$. Against uniform sampling of permutations it is better for assignment ($0.114$ versus $0.042$) and on par for QAP ($0.063$ versus $0.058$) and TSP ($0.330$ versus $0.333$); the quality inside the feasible set is only modestly above uniform. With only 24 feasible permutations (and 8 optimal tours for TSP) these instances are small, and the comparison with uniform sampling is the relevant one.

\section{Discussion}
\label{sec:discussion}
\paragraph{What is gained, and from where.}
The improvement over penalty QAOA in Sec.~\ref{sec:results} does not come from a better circuit in a narrow sense:
by the Corollary of Sec.~\ref{sec:theory}, Two-Step QAOA has exactly the post-selected landscape of the Dicke-initialized
alternating-operator ansatz, and indeed matches it up to the factor $P_{\mathcal G}$ whenever training reaches the same angles. What Two-Step QAOA adds is a way to obtain that
ansatz from the penalty formulation the user already has, with no penalty weight to search and no state-preparation
circuit to design. The comparison that matters for a user is therefore against penalty QAOA \emph{with a realistic way of
choosing the weight}, which is what our baselines implement.

\paragraph{Penalty QAOA can be competitive if the weight is known.}
When the weight is chosen with oracle knowledge (the exact optimum, used to pick the best of five weights) and five times
the training budget is spent, penalty QAOA overtakes ring-mixer Two-Step at larger depth for half filling, and a ladder started
from a small weight does the same; complete-graph Two-Step stays ahead of both.
We read this as evidence that the penalty weight is the crux, not that penalties are useless: Two-Step QAOA is useful
precisely because the good weight is not known in advance.

\paragraph{Cheaper initial states.}
The biased product state $\bigotimes_i(\sqrt{1-k/n}\,|0\rangle+\sqrt{k/n}\,|1\rangle)$ needs no entangling gate and, by the
same symmetry argument as Proposition~2, its weight-$k$ component is exactly the Dicke state. With post-selection it gives
exactly the same landscape, at the price of a feasible fraction $(1-1/n)^{n-1}\to1/e$ for $k=1$ and $\binom{n}{n/2}2^{-n}$
for half filling ($k=n/2$), the same as $|+\rangle^{\otimes n}$. Stage~1 trades entangling layers for shots: each of its layers costs $n(n-1)$ CNOTs on all-to-all
hardware, and it raises the kept fraction of samples from $P_k^{\rm product}$ to $P_{\mathcal G}$, i.e.\ it saves $1/P_k^{\rm product}-1/P_{\mathcal G}$ shots per kept sample. Which side of this trade is better depends
on the hardware (native XY or ZZ interactions, connectivity, error rates), which is why we do not use CNOT counts as the
figure of merit; on hardware with cheap two-qubit gates stage~1 is preferable, on hardware where shots are cheap and gates
are not, the biased product state is.

\paragraph{The mixer matters.}
With the ring mixer, a small penalty weight overtakes Two-Step at depth three to five for half filling; with the complete-graph mixer it does not. Since the post-selected landscape of Two-Step equals that of the Dicke-initialized ansatz, everything known about XY-mixer design~\cite{wang2020xy} carries over, and the choice of edge set should be part of the compilation pass.

\paragraph{Permutation constraints.}
Assignment, QAP and TSP impose one-hot constraints on both rows and columns. Only one family can be hardened with
group-wise XY mixers, and the other must stay a penalty. In our $4\times4$ instances this raised the feasible fraction
from a few percent to $0.6$--$0.85$ and improved on penalty QAOA by a factor $14$--$34$, but the quality inside the feasible set
stayed close to uniform, so uniform sampling of permutations is as good for QAP and TSP. These problems need mixers that
preserve permutations~\cite{Hadfield_2019}; Two-Step QAOA's stage~1 does not provide the corresponding initial state.

\paragraph{Limitations.}
All results are noiseless state-vector simulations with up to 16 qubits and small instance sets (5--20 per problem) (stage-1 tables up to 32 qubits per group).
Training uses exact expectations and automatic differentiation; the shot counts we report are a budget proxy
(two parameter-shift evaluations per angle) and not a finite-shot optimization. Noise that changes group weights (e.g.\ amplitude damping or bit flips; dephasing in the computational basis does not) breaks the conservation law of
Proposition~1, so on hardware the post-selection also filters some noise-induced errors and its overhead grows in a noise-model-dependent way; we have
not quantified this. The stage-1 table entries are the best values found by a local optimizer and are lower bounds on what
is achievable. Finally, the matrix-based detector is heuristic; constraints should be passed explicitly.

\section{Conclusion}
Two-Step QAOA turns the k-hot penalties a user already writes into hard constraints, without a penalty weight and without a dedicated Dicke-state circuit.
Its stage 1 is an exact Dicke-state preparation up to leakage, with problem-independent angles that can be shipped as a table, and its stage 2 has the landscape of the Dicke-initialized alternating-operator ansatz.
In our simulations of portfolio problems this removes the penalty search and, with a complete-graph XY mixer, gives a higher median success probability than every penalty-selection rule we tested at every depth up to five; with the ring mixer the advantage is limited to shallow depth. For graph coloring, where all constraints are hardened (ring mixer, compared with the formulation library's penalty ladder and uniform sampling), it finds optimal colorings far more often. For permutation constraints, where only part of the constraints can be hardened, it improves on penalty QAOA but not clearly on uniform sampling, and permutation-preserving mixers are the natural next step.

\section*{Code and data availability}
The simulations use the blueqat SDK (GitHub version 2.2.0) for circuit construction and verification. The scripts, the tables of stage-1 angles and the raw results behind all numbers in this paper are available at \url{https://github.com/minatoyuichiro/two-step-qaoa}. The detection and multi-group experiments additionally use an in-house library of QUBO formulations that is not public; their raw results are included in the repository.

\section*{Acknowledgments}
I would like to express my gratitude to Takao Tomono from Keio University and independent researcher Shunsuke Sotobayashi for their invaluable contributions and support.

\bibliographystyle{unsrt}
\bibliography{references}

@misc{farhi2014quantumapproximateoptimizationalgorithm,
      title={A Quantum Approximate Optimization Algorithm}, 
      author={Edward Farhi and Jeffrey Goldstone and Sam Gutmann},
      year={2014},
      eprint={1411.4028},
      archivePrefix={arXiv},
      primaryClass={quant-ph},
      url={https://arxiv.org/abs/1411.4028}, 
}

@article{niroula2022constrained,
  title={Constrained quantum optimization for extractive summarization on a trapped-ion quantum computer},
  author={Niroula, Pradeep and Shaydulin, Ruslan and Yalovetzky, Romina and others},
  journal={Scientific Reports},
  volume={12},
  number={1},
  pages={17171},
  year={2022},
  publisher={Nature Publishing Group},
  doi={10.1038/s41598-022-20853-w}
}

@inproceedings{Tabi_2020,
   title={Quantum Optimization for the Graph Coloring Problem with Space-Efficient Embedding},
   url={http://dx.doi.org/10.1109/QCE49297.2020.00018},
   DOI={10.1109/qce49297.2020.00018},
   booktitle={2020 IEEE International Conference on Quantum Computing and Engineering (QCE)},
   publisher={IEEE},
   author={Tabi, Zsolt and El-Safty, Kareem H. and Kallus, Zsofia and Haga, Peter and Kozsik, Tamas and Glos, Adam and Zimboras, Zoltan},
   year={2020},
   month=oct }

@article{Hadfield_2019,
   title={From the Quantum Approximate Optimization Algorithm to a Quantum Alternating Operator Ansatz},
   volume={12},
   ISSN={1999-4893},
   url={http://dx.doi.org/10.3390/a12020034},
   DOI={10.3390/a12020034},
   number={2},
   journal={Algorithms},
   publisher={MDPI AG},
   author={Hadfield, Stuart and Wang, Zhihui and O’Gorman, Bryan and Rieffel, Eleanor G. and Venturelli, Davide and Biswas, Rupak},
   year={2019},
   month=feb, pages={34} }

@inproceedings{B_rtschi_2019,
   title={Deterministic Preparation of Dicke States},
   ISBN={9783030250270},
   ISSN={1611-3349},
   url={http://dx.doi.org/10.1007/978-3-030-25027-0_9},
   DOI={10.1007/978-3-030-25027-0_9},
   booktitle={Fundamentals of Computation Theory (FCT 2019)},
   pages={126--139},
   publisher={Springer International Publishing},
   author={Bärtschi, Andreas and Eidenbenz, Stephan},
   year={2019},
 }

@article{Suksmono_2022,
   title={Quantum computing formulation of some classical Hadamard matrix searching methods and its implementation on a quantum computer},
   volume={12},
   ISSN={2045-2322},
   url={https://doi.org/10.1038/s41598-021-03586-0},
   DOI={10.1038/s41598-021-03586-0},
   number={197},
   journal={Scientific Reports},
   publisher={Springer Nature},
   author={Suksmono, A. B. and Minato, Y.},
   year={2022}
}

@article{wang2020xy,
   title={{XY}-mixers: Analytical and numerical results for the quantum alternating operator ansatz},
   author={Wang, Zhihui and Rubin, Nicholas C. and Dominy, Jason M. and Rieffel, Eleanor G.},
   journal={Physical Review A},
   volume={101},
   pages={012320},
   year={2020},
   doi={10.1103/PhysRevA.101.012320}
}

@inproceedings{baertschi2022short,
   title={Short-Depth Circuits for {Dicke} State Preparation},
   author={B{\"a}rtschi, Andreas and Eidenbenz, Stephan},
   booktitle={2022 IEEE International Conference on Quantum Computing and Engineering (QCE)},
   year={2022},
   doi={10.1109/QCE53715.2022.00027}
}

@article{zhou2020qaoa,
   title={Quantum Approximate Optimization Algorithm: Performance, Mechanism, and Implementation on Near-Term Devices},
   author={Zhou, Leo and Wang, Sheng-Tao and Choi, Soonwon and Pichler, Hannes and Lukin, Mikhail D.},
   journal={Physical Review X},
   volume={10},
   pages={021067},
   year={2020},
   doi={10.1103/PhysRevX.10.021067}
}

@article{lucas2014ising,
   title={Ising formulations of many {NP} problems},
   author={Lucas, Andrew},
   journal={Frontiers in Physics},
   volume={2},
   pages={5},
   year={2014},
   doi={10.3389/fphy.2014.00005}
}

\end{document}